\documentclass[12pt, preprint]{aastex}
\usepackage{natbib}

\def\be{\begin{equation}}
\def\ee{\end{equation}}
\def\lesssim{\raisebox{-0.3ex}{\mbox{$\stackrel{<}{_\sim} \,$}}}
\def\gtrsim{\raisebox{-0.3ex}{\mbox{$\stackrel{>}{_\sim} \,$}}}
\def\degsp{\hbox{$^{\circ}$}}

\begin{document}
\title{Curvature radiation and giant subpulses in the Crab pulsar}
\author{Janusz Gil\altaffilmark{1} \& George I. Melikidze\altaffilmark{1,2}}
%\email{jag@astro.ia.uz.zgora.pl}
\altaffiltext{1}{Institute of Astronomy, University of Zielona
G\'ora, Lubuska 2, 65-265, Zielona G\'ora, Poland}
\altaffiltext{2}{Abastumani Astrophysical Observatory, Al.Kazbegi
ave. 2a, Tbilisi 0160, Georgia }

\begin{abstract}
It is argued that the nanosecond giant subpulses detected recently
in the Crab pulsar are generated by means of the coherent
curvature radiation of charged relativistic solitons associated
with sparking discharges of the inner gap potential drop above the
polar cap.
\end{abstract}

\keywords{pulsars: giant pulses - pulsars: individual (Crab pulsar)}
\section{Introduction}

Although only four pulsars are known to emit giant pulses
\citep{lun95,cog96,rj01,jr03}, understanding the mechanism of
their radiation can potentially lead to understanding the
longstanding problem of a pulsar radio emission. Recently, the
detection of extremely short and powerful, 2-nanosecond - 1000 Jy,
subpulses within the radio giant pulses from the Crab pulsar has
been reported by \citet[][ HKWE03 hereafter]{hetal03}. This is an
observational result of extraordinary importance, for it can shed
a new light onto the mystery od coherent pulsar radio emission.
HKWE03 argued that these nanosecond giant subpulses were true
temporal modulations associated with an explosive collapse of
nonlinear plasma turbulence \citep{wetal98}. In this letter we
argue that the time-scale of the Crab pulsar giant nanosecond
subpulses is consistent with the angular beaming due to the
curvature of dipolar field lines (different from the conventional
angular beaming due to the pulsar rotation), thus with the
coherent curvature radiation. Furthermore, we demonstrate that the
observed fluxes of these giant subpulses are consistent with the
emission of charged relativistic solitons, generated by means of
the modulational instability of the strong Langmuir turbulence
associated with a sparking discharge of the pulsar's polar gap
(Melikidze, Gil \& Pataraya 2000, MGP00 hereafter; Gil, Lyubarski
\& Melikidze 2004, GLM04 hereafter).

\section{The time-scales}
\subsection{The Alignment Time-scale}

Let us consider a number of localized (point-like) sources of the
broadband coherent curvature radiation, moving relativistically
(with the Lorentz factor $\gamma=(1-\beta^2)^{-1/2}$, where
$\beta=v/c\sim 1$) along a narrow bundle of dipolar magnetic field
lines with a radius of curvature \be \rho=1.8 r_6^{1/2}s^{-1}10^7
~{\rm cm}, \label{ro} \ee where $s=d/r_p$ is the normalized polar
coordinate of the central line of the bundle, $r_6=r_{\rm em}/R$
is the normalized emission altitude $r$ (for the frequency
$\nu=5~{\rm GHz},~5<r_6<35$ in the Crab pulsar, Kijak \& Gil
1998), $d$ is the distance from the dipolar axis to the foot of a
dipolar field line, $r_p\approx 10^4P^{-1/2}$ is the polar cap
radius and $R=10^6$~cm is the neutron star radius. The geometry is
schematically illustrated in Fig.~1 where a narrow bundle of field
lines is marked. The curvature radiation is emitted along dipolar
field lines into a narrow cone with an opening angle
$\theta=1/\gamma$ \citep[e.g.][]{j75}. In what follows we assume
that the perpendicular dimension of each source measured by the
angular extent of the bundle of field lines does not exceed
$\theta$ (validity of this assumption is justified later on in the
paper). We also assume that other bundles carrying sources of
coherent curvature radiation are separated by at least $\theta$,
thus only one narrow bundle contributes to the radiation observed
at a given rotational phase $\varphi=2\pi t/P$, where $t$ is the
intrapulse time. For a fixed arrangement of the observer's
direction, the field line polar coordinate $s$ and the frequency
dependent emission altitude $r_6(\nu)$, each source is aligned
within $1/\gamma$ with the observer during the time interval
$\delta t=l/v=\rho\theta/v=\rho\gamma^{-1}/v$ (see Fig.~1), where
$\rho$ is described by equation (\ref{ro}). Therefore, a
continuous stream of relativistic sources moving along a curved
trajectory would illuminate the observer during the time interval
\begin{figure}
\epsscale{1}\plotone{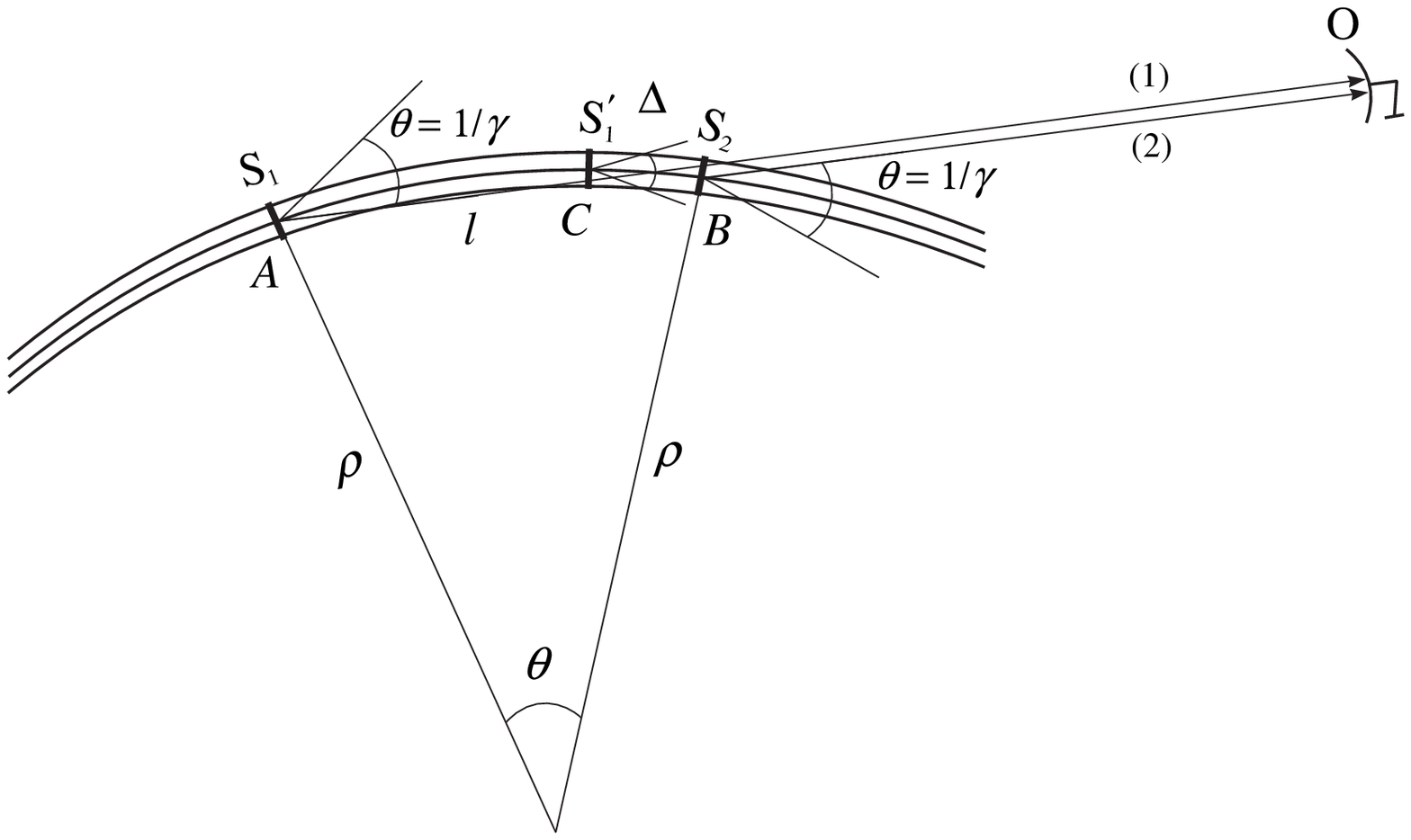} \caption{Geometry of the observed
curvature radiation emitted by the source ($S$) moving
relativistically along a narrow bundle od dipolar field lines with
the radius of curvature $\rho$ (see text for explanation of other
symbols).}
\end{figure}
\be \Delta t=\rho\gamma^{-1}/c~, \label{ti} \ee provided that this
``alignment'' time-scale is longer than the conventional angular
beaming time scale $\Delta t'=\gamma^{-1}P/2\pi$ related purely to
the pulsar rotation. In the case of the Crab pulsar this means
that $\rho>1.6\times 10^8$~cm in the case of the Crab pulsar.
HKWE03 concluded that the conventional angular beaming cannot
explain the 2-nanosecond duration of their giant subpulses, since
it indeed requires $\gamma\sim 10^6$, much above the expected
values $\gamma<1000$. Since, incidentally, in the case of the Crab
pulsar $\Delta t'\sim\Delta t$, their conclusion applies also to
the alignment time scale related to the field line curvature
(eq.[\ref{ti}]).

\subsection{The Apparent Time-Scale}

The situation can be quite different if the stream is not
continuous but some sources are distinguished by emitting slightly
more power ${\cal P}$ than the average. From the observational
point of view it means that there are very few sources emitting
towards the observer. For simplicity let us first consider a
single source, which is marked in Fig.~1 in three different
positions A, B and C. Following Jackson (1975) we can argue that
the duration of the observed impulse related to the curvature of
field lines is much shorter than that predicted by equation
(\ref{ti}). Let us consider two particular points along a given
dipolar field line: first (A) at which the source $S_1$ becomes
aligned (ray 1) and second (B) at which it becomes misaligned (ray
2) with the observer $O$ (Fig.1). The source covers a distance
$l=\rho/\gamma$ between these points during a time interval
$\delta t=\rho\gamma^{-1}/v$. During this time the radiation
emitted at the first alignment point (A) travels a distance
$L=c\delta t=(c/v)\rho\gamma^{-1}=\rho/(\beta\gamma)$. Thus, the
radiation overtakes the source only by a distance
$\Delta=L-l=(1/\beta-1)(\rho/\gamma)\approx\rho/\gamma^3$ (when
the radiation emitted at point A reaches point B, the source of
this radiation arrives at point C). This is the pulse length in
space, and thus the duration of the observed impulse emitted by
the considered source is \be
\Delta\tau=\frac{\Delta}{c}=\frac{\rho}{c\gamma^3}=\gamma^{-2}\Delta
t~,\label{dtau} \ee (Gil 1985) where $\Delta t$ is the alignment
time scale (eq.[\ref{ti}]). Therefore, if the longitudinal (along
field lines) source dimension is smaller than
$\Delta\approx\rho/\gamma^3$, then the apparent duration of the
impulse $\Delta\tau$ is $\gamma^2$ times shorter than the actual
time of alignment of the source with the observer. It is important
to emphasize that this apparent shortening effect has nothing to
do with special relativity (time dilation) and/or Doppler effects,
as both $\Delta t$ and $\Delta\tau$ are referred to the same
observer's frame. This effect occurs for any localized source of
emission (radiation, particles, etc.) moving towards the observer
(target) with a speed slightly lower that the emitted waves
(particles). For example, one can consider an acoustic analog in
which a highly directional loud-speaker is moving along a curved
trajectory with a velocity close to the speed of sound. The
duration of the acoustic impulse is much shorter than the actual
time of geometrical alignment of the beam pattern with the
recording device. The acoustic analog of equation (\ref{dtau}) is
$\Delta\tau=\Theta^2\Delta t$, where $\Theta$ is the beam-width of
the emitted pattern (in radians).

Both isolated 2-ns giant subpulses, as well as a sequence of a
number of such subpulses (when unresolved this sequence
constitutes a normal giant pulse in our view), can be seen in
Fig.~1 of HKWE03. Within our model such case corresponds to a
sequence of sources, each having a longitudinal dimension smaller
than $\Delta$ and separation between the adjacent sources larger
than $\Delta$, where $\Delta\approx\rho/\gamma^3$ is the spatial
length of the impulse associated with each source. For each giant
subpulse detected by HKWE03 the observed time duration
$\Delta\tau=2\times 10^{-9}$~s (eq.[\ref{dtau}]), which leads to
the condition \be \rho_8\approx 0.6\gamma_2^3~,\label{ro8} \ee
where $\rho_8=\rho/10^8$~cm (eq.[\ref{ro}]) and
$\gamma_2=\gamma/10^2$.

\section{Energetics}
\subsection{Brightness Temperature}
As reported by HKWE03, the ultra-short nano-giant subpulses
detected at $\nu=5$~GHz in the Crab pulsar often exceed fluxes
$S=1000~{\rm Jy}=10^{-20}{\rm erg~s}^{-1}{\rm Hz}^{-1}{\rm
cm}^{-1}$. These fluxes can be converted into the brightness
temperature $T_b=SD^2/[2k(\nu W)^2]$, where $D=2~{\rm kpc}=6\times
10^{21}{\rm cm}$ is the distance to the pulsar, $k$ is the
Boltzmann constant and $W$ is the time scale of the corresponding
emission process \citep[e.g.][]{mc03}. Assuming that $W=2\times
10^{-9}$s (HKWE03) one obtains the extraordinarily high brightness
temperatures $T_b\sim 10^{38}$K, implying by far the most luminous
emission from any astronomical object. However, within our model
the actual emission process corresponding to a giant subpulse of
duration $\Delta\tau=2\times 10^{-9}$s occurs over a much longer
``alignment'' time interval $W=\Delta t=\gamma^2\Delta\tau$
(eqs.~[\ref{ti}] and [\ref{dtau}]), which leads to $\gamma^4$
times lower brightness temperatures. Since $\gamma$ is of the
order of 100, then $T_b<10^{30}$~K, consistent with normal giant
pulses in the Crab pulsar and other pulsars (see Fig.~1 in
McLaughlin and Cordes, 2003).

\subsection{Luminosity and Power}
Let us then consider a source(s) moving along a dipolar field line
(Fig.~1) and emitting a coherent curvature radiation with an
intrinsic power ${\cal P}$. While moving over the alignment
distance $l=\rho\gamma^{-1}$, the emitted energy $E={\cal P}\Delta
t$, where the alignment time interval $\Delta t=\rho\gamma^{-1}/c$
(eq.[\ref{ti}]). This energy is received by the observer in a much
shorter time $\Delta\tau=\gamma^{-2}\Delta t$ (eq.[\ref{dtau}]),
and therefore the apparent luminosity \be {\cal
L}=\frac{E}{\Delta\tau}=\gamma^2{\cal P}. \label{lum} \ee Thus,
the nanosecond giant subpulses appear $\gamma^2=\gamma^2_210^4$
more luminous than the intrinsic power of their source(s). Also
the apparent fluxes $S\propto{\cal L}$ are $\gamma^2$ times
overestimated. It is important to emphasize that this kinematical
boosting is a direct consequence of an apparent shortening of the
impulse duration described by equation (\ref{dtau}) and again has
nothing to do with special relativity and/or Doppler effects
\citep{gin79}. According to the discussion below equation (\ref{dtau}),
an acoustic analog of equation (\ref{lum}) is ${\cal L}=\Theta^{-2}{\cal
P}$.

The apparent fluxes $S\sim 1000$~Jy of the nanosecond giant
subpulses can be formally converted into the emitted luminosity
${\cal L}=S\theta^2D^2\Delta\nu$. With $D=2$~kpc,
$\theta=\gamma_2^{-1}10^{-2}$ and $\Delta\nu\sim 10^9$~Hz one
obtains ${\cal L}\sim\gamma_2^{-2}10^{28}$~erg/s. This is an
extremely large value, comparable with the total radio luminosity
of the Crab pulsar $L_R=3.5\times 10^{25+x}$~erg~s$^{-1}$, where
$x=logL\approx 3.6$ mJy~kpc$^2$ from the Pulsar Catalog
\citep{tml93}. The existence of such powerful localized sources of
coherent radio emission seems highly questionable (of course, this
argument is not independent from the extraordinarily high
brightness temperatures $T_b\sim 10^{38}$K derived in
Section~3.1). However, according to equation (\ref{lum}), the
actual power of the emitted radiation ${\cal
P}=\gamma^{-2}S\theta^2D^2\Delta\nu=\gamma^{-2}{\cal L}$, which
for the parameters given above yields approximately \be {\cal
P}\sim \gamma_2^{-4}10^{24}~{\rm erg~s}^{-1}. \label{p} \ee This
is still a very large power (consistent with $T_b\lesssim
10^{30}$~K) but much smaller than the total radio luminosity of
the Crab pulsar. In the next section we discuss a mechanism of
coherence of the curvature radiation that can produce shots of
radio emission of nanosecond duration with the intrinsic power
corresponding to the values described by equation (\ref{p}).

Finally we can check whether our localized source can be supplied
with enough kinetic energy to emit giant subpulses. To estimate
the energy of the source let us note that the product of plasma
number density and the cross-section of the corresponding flux
tube is a constant value equal to $\kappa n_{\rm GJ} h^2$, where
$n_{\rm GJ}$ is the \citet{gj69} density, $\kappa$ is the
\citet{stu71} multiplication factor and $h$ is the gap height
(equal to the spark size). Assuming that the perpendicular source
dimensions are determined by the spark size, we can write that the
kinetic energy associated with the source is $E_{\rm sr}\approx
n_{\rm GJ}\kappa m c^3h^2{\Delta}/{c}$. On the other hand, the
giant subpulse energy $E_{\rm GP}=\gamma^2 {\cal P}
\Delta\tau=\gamma^2 {\cal P} \Delta/c$ (eqs.[3] and [5]).
Obviously $E_{\rm GP}$ should be smaller than $E_{\rm sr}$, which
leads to the condition $\kappa\gamma_{2}^3>500$, consistent with
an independent estimate given by GLM04.

\section{Coherent curvature radiation}

MGP00 proposed a model for generation of the coherent pulsar radio
emission, based on the idea of the polar gap discharging via a
number of localized sparks \citep{rs75}. The sparking phenomenon
creates  a succession of plasma clouds moving along dipolar
magnetic field lines. The overlapping of charged particles with
different energies from the adjacent clouds ignites a strong
Langmuir turbulence via the two-stream instability
\citep{usov87,am98}. In the pulsar magnetospheric conditions, this
turbulence is subject to modulational instability, which leads to
the formation of ``bunch-like'' charged solitons capable of
generating the coherent curvature emission at radio-wavelengths
$\lambda$. Thus, the condition
$\lambda<\Delta=\rho/\gamma^3=(\rho_8/\gamma_2^3)10^2$ is
naturally satisfied at centimeter wavelengths (note that according
to eq.[\ref{ro8}] $\Delta =60$~cm, which of course corresponds to
2 light nanoseconds). Additionally, the angular extent
$\Delta\psi$ of the bundle of dipolar field lines associated with
a given spark should not exceed $\theta=\gamma^{-1}$. It is easy
to show that in the Crab pulsar $\Delta\psi\sim 0.001<\gamma^{-1}$
\citep[for details see][]{gs00}.

MGP00 calculated the power of the soliton curvature radiation in
the vacuum approximation, that is without taking into account the
influence of the magnetospheric plasma. GLM04 reconsidered the
curvature radiation of a point-like charge moving through
electron-positron plasma. They demonstrated that the radiation
power is suppressed by a factor of about $10^{-2}-10^{-3}$ but
still at a high enough level to explain the observed pulsar
luminocities. Applying the results of MGP00 modified by GLM04 to
the case of the Crab pulsar, we obtain that the power radiated by
one soliton can be as high as $L_1\sim 10^{21}$~erg/s. Thus, for
an incoherent superposition of $N$ sources of the coherent
radio-emission (solitons), each smaller than $\Delta$ and
separated from each other by more than $\Delta$, we have \be L\sim
N10^{21}~{\rm erg/s},\label{L} \ee (for details see eqs.~[14] -
[18] in MPG00). Comparing equation (\ref{L}) with the power required for
the Crab giant subpulses expressed by equation (\ref{p}) one obtains the
Lorentz factors $\gamma_2\gtrsim 10^{3/4}/N^{1/4}$, which is
realistically about 3-4 (consistent with an independent estimate
for $\gamma\sim 300-400$ given by GLM04). Inserting $\gamma_2=3$
into equation (\ref{ro8}) we obtain $\rho_8\gtrsim 16$, which, according
to equation (\ref{ro}), means $r_6^{1/2}s^{-1}\sim 90$. Since $r_6<35$
\citep{kg98} this implies that $s<0.07$. Remembering that $s$ is a
normalized polar coordinate (0 - for the pole and 1 for the polar
cap edge), we conclude that only those field lines that originate
very close to the magnetic axis can be involved with the
nanosecond giant sub-pulses. Such field lines can be aligned with
the observer in the radio emission region only within a very
narrow range of longitudes near the so-called fiducial plane,
containing both the spin and magnetic axes. It is natural to
associate this plane with the peak of the pulse profile. Thus,
within our model, the giant subpulses should occur only within a
very narrow range of longitudes near the peak of the radio
profile. To the best of our knowledge, giant subpulses in the Crab
pulsar seem to occupy the narrow range of phases within about 1\%
of the pulse window near the peak of the main-pulse (HKWE03). This
is consistent with the phase range
$2\psi/360\degsp=s(2.\degsp4/360\degsp)(r_6/P)^{1/2}$, which for
$s<0.07$ and $r_6<35$ (see above) is smaller than 0.015.

\section{Discussion}

In this letter we argue that the 2-ns giant subpulses detected in
the Crab pulsar are due to the curvature radiation of one or at
most several solitons associated with a sparking discharges
occurring near the local surface magnetic pole. This is the most
elementary emission event that can be observationally resolved.
The curvature radiation from the particular soliton(s) can be
observationally distinguished by means of the kinematical boosting
(eq.[\ref{dtau}]) only if they are slightly more powerful than
other solitons. This implies that the sparking event associated
with giant subpulses is more energetic than an average one. Both
luminosity and characteristic frequency of the soliton curvature
radiation strongly depend on the Lorentz factor $\gamma$, which in
turn depends on the accelerating potential drop above the polar
cap. Interestingly, pulsars exhibiting giant pulses are
distinguished not only by their brightness, but also by extremely
high values of the so-called complexity parameter $a=(r_p/h)$,
where $r_p$ is the polar cap radius and $h$ is the polar gap
height \citep{gs00}. Since the accelerating potential drop $\Delta
V\propto h^2$ and $h<r_p/\sqrt{2}$ \citep{rs75}, then pulsars with
very high values of $a>>1$ must have a large reservoir of the
maximum available potential drop over the actual potential drop
$(\Delta V_{max}/\Delta V=a^2/2)$. This reservoir can be
occasionally used to create exceptionally energetic spark(s).

\begin{deluxetable}{lccccc}
\tabletypesize{\scriptsize} \tablecaption{Pulsar
parameters\label{table1}} \tablecomments{Pulsars marked in
boldface show giant pulses, others are candidates for giant pulse
emission. Numbers in parenthesis denote ranking position in
$B_{LC}=2850\dot{P}^{0.5}P^{-2.5}$. The complexity parameter
$a=425(\dot{P})^{2/7}P^{-9/14}$.} \tablewidth{3in}
\tablehead{\colhead{PSR J} &
\colhead{$P$} & \colhead{$\dot{P}$} & \colhead{$a$} &  \multicolumn{2}{c}{$B_{LC}$} \\
& \colhead{(msec)} & \colhead{$(10^{-15})$} & &
\multicolumn{2}{c}{($10^5$G)}}

\startdata {\bf 0534+2200} & 34 & 420 & 249 & 8.7 & (2) \\
{\bf 0540$-$6919} & 51 & 479 & 198 & 3.3 & (6) \\
1513$-$5908 & 150 & 1540 & 138 & 0.4 & (21) \\
1124$-$5916 & 135 & 745 & 120 & 0.4 & (22) \\
1617$-$5055 & 69 & 137 & 114 & 0.8 & (12) \\
1420$-$6048 & 68 & 83 & 99 & 0.7 & (14)\\
1119$-$6127 & 408 & 4002 & 95 & 0.05 & (75)\\
{\bf 0835$-$4510} & 89 & 125 & 94 & 0.4 & (20)\\
{\bf 1824$-$2452} & 3 & 0.0016 & 33 & 7.3 & (4)\\
1823$-$30A & 5.4 & 0.0034 & 28 & 2.4 & (8)\\
{\bf 1939$+$2134} & 1.6 & 0.0001 & 23 & 8.8 & (1)\\
{\bf 0218$+$4232} & 2.3 & 0.00007 & 16 & 3.1 & (7)\\
{\bf 1959$+$2048} & 1.6 & 0.00002 & 14 & 3.6 & (5)\\
2129$+$1210E & 4.6 & 0.00018 & 14 & 8.2 & (3)\\
\enddata
\end{deluxetable}

The values of the complexity parameter $a$ can well exceed 100 in
normal pulsars, while in millisecond pulsars they are limited to
about 40 \citep[see Fig.~1 in][]{gs00}. Therefore, when pulsars in
both groups are ranked with respect to $a$, pulsars with giant
pulses occupy the top of the lists (Table 1). Also, the Vela
pulsar, which was reported to exhibit some kind of giant pulse
behavior \citep{joh01} is very high on the list. Moreover,
sporadic large amplitude pulses (LAP) from two millisecond pulsars
(J1959+2048 and J0218+4232) have been reported very recently by
\citet{jos03}. It should be emphasized that pulsars showing giant
pulses have also high values of the magnetic field $B_{\rm
LC}\gtrsim 10^5$~G at the light cylinder. This might, however, be
a coincidence because roughly $B_{\rm LC}\sim a^2/P$, where $P$ is
the pulsar period (Table 1). Although $B_{\rm LC}$ is a good
parameter to make a list of giant pulse candidates, only a value
of $a^2$ has a physical meaning within our model. In Table 1 we
propose a number of candidates for giant pulses with high values
of the complexity parameter ($a>10$ for millisecond pulsars and
$a>90$ for normal pulsars) and relatively low values of $B_{\rm
LC}$. The detection (non-detection) of giant pulses from these
candidates would confirm (refute) our scenario. PSR J1119-6127
with low $B_{\rm LC}\approx 5\times 10^3$ G seems the most
interesting case.

The case of PSR J1959+2048 deserves some additional discussion in
the light of the model discussed here. As emphasized by
\citet{jr03}, this pulsar is exceptional in the sense that despite
the high value of $B_{\rm LC}$ it does not show giant emission
(although \citet{jos03} found one significant LAP (out of million
pulses) with an intensity $129$ times the mean intensity). This
seems to be consistent with our model since this pulsar (which is
fourth ranked with respect to $B_{\rm LC }$) has a relatively
small value of the complexity parameter $a$ as compared with two
other millisecond pulsars showing giant pulses (Table 1). In fact,
since $\Delta V_{\rm max}/\Delta V=a^2/2$, the smaller the value
$a$, the smaller the reservoir of the polar gap energy that can
ignite giant emission.

Also the case of Vela pulsar is very interesting in view of the
models of giant emission. \citet{kjv02} reported giant
micro-pulses in this pulsar, which are different from normal
micro-pulses and are probably closely related to classical giant
pulses. They have typically large amplitudes, appear to be
narrower than normal micro-pulses and exhibit a power law in their
cumulative probability distribution. The first two properties are
quite natural within our model (a sequence of unresolved
nano-pulses?) and we discuss the last issue bellow.

According to our scenario, both ordinary and giant pulses are
related to the inner gap sparking activity and originate at
relatively low altitudes, contrary to the suggestion that the
latter arise in the outer gap region \citep{rj01}. Generally, the
following two steps are involved: first a corresponding spark
should be relatively intense (which corresponds to relatively high
Lorentz factors), so the associated solitons are distinguished
from the background radiation (the soliton radiation intensity
depends very strongly on the Lorentz factor, see eq.[43] in
GLM04). Then, in the second step, the intensity of the curvature
radiation of distinguished soliton(s) can be kinematically boosted
by means of equation (\ref{lum}). It is intuitive that the
distribution of spark energy is quasi-gaussian, with a low energy
cutoff corresponding to the pair formation threshold. Since the
giant pulses should be associated with sparks corresponding to the
high energy tail of this distribution, it is natural that they
exhibit a power law in their cumulative probability distribution.

\acknowledgments This paper is supported in part by the Grant
2~P03D~008~19 of the Polish State Committee for Scientific
Research. We thank D. Melrose and Q. Luo for their hospitality and
fruitful discussions during our stay at the School of Physics
Sydney University. We thank E. Gil and U. Maciejewska for
technical help.

{}


\begin{thebibliography}{}

\bibitem[Asseo \& Melikidze(1998)]{am98}  Asseo, E., \& Melikidze, G. I. 1998, \mnras, 301, 59
\bibitem[Cognard et al.(1996)]{cog96} Cognard, L., Shrauner, J.A., Taylor, J.H. \& Thorsett, S.E., 1996, \apj, 457, L81
\bibitem[Gil(1985)]{g85} Gil, J., 1985, A\&A 143, 443-446
\bibitem[Gil, Lyubarsky \& Melikidze(2004)]{glm04} Gil, J., Lyubarsky, Y., \& Melikidze, G.I.
2004, \apj, in press (GLM04), astro-ph/0310621
\bibitem[Gil \& Sendyk(2000)]{gs00} Gil, J. \& Sendyk, M. 2000,
\apj, 541, 351
\bibitem[Ginzburg(1979)]{gin79} Ginzburg, V.L. 1979, Theoretical physics and Astrophysics, sec. 5, Pergamon Press, Oxford
\bibitem[Goldreich \& Julian(1969)]{gj69}  Goldreich, P., \& Julian, H. 1969, \apj, 157, 869
\bibitem[Hankins et al.(2003)]{hetal03} Hankins, T.H., Kern, J.S., Weatherall, J.C. \& Eilek, J.A.
2003, Nature, 422, 141 (HKWE03)
\bibitem[Jackson(1975)]{j75} Jackson, J. D. 1975, Classical electrodynamics, John Wiley \& Sons, NY
\bibitem[Johnston \& Romani(2003)]{jr03} Johnston, S., \& Romani, R.W. 2003, astro-ph/0305235
\bibitem[Johnston et al.(2001)]{joh01} Johnston, S., et al. 2001, \apj, 549, L101
\bibitem[Joshi et al.(2003)]{jos03} Joshi, B.C. et al. 2003, astro-ph/0301285
\bibitem[Kramer, Johnston \& van Straten(2002)]{kjv02} Kramer, M., Johnston, S., \& van Straten, W. 2002, \mnras. 334, 523
\bibitem[Kijak \& Gil(1998)]{kg98}  Kijak, J., \& Gil, J. 1998, \mnras, 299, 855
\bibitem[Lundgren et a.(1995)]{lun95} Lundgren, S.C., et al., \apj, 453, 433
\bibitem[McLaughlin and Cordes(2003)]{mc03} McLaughlin, M.A., \&
Cordes, J.M. 2003, astro-ph/0304365
%\bibitem[Melikidze \& Pataraya(1980)]{mp80}  Melikidze, G. I., \& Pataraya, A. D. 1980, Astrofizika, 16, 161
%\bibitem[Melikidze \& Pataraya(1984)]{mp84}  Melikidze, G. I., \& Pataraya, A. D. 1984, Astrofizika, 20, 157
\bibitem[Melikidze, Gil \& Pataraya(2000)]{mgp00} Melikidze, G. I, Gil, J., \& Pataraya, A. D. 2000, ApJ, 544,
1081 (MGP00)
\bibitem[Romani \& Johnston(2001)]{rj01} Romani, R.W., \& Johnston, S. 2001, \apj, 557, L93
\bibitem[Ruderman \& Sutherland(1975)]{rs75} Ruderman, M. A., \& Sutherland, P. G. 1975, \apj, 196, 51
\bibitem[Sturrock(1971)]{stu71} Sturrock, P. A. 1971, \apj, 164, 529
\bibitem[Taylor, Manchester \& Lyne(1993)]{tml93} Taylor, J.H.,
Manchester, R.N., \& Lyne, A.G. 1993, ApJS, 88, 259
\bibitem[Usov(1987)]{usov87} Usov, V. V. 1987, \apj, 320, 333
\bibitem[Weatherall et al.(1998)]{wetal98} Weatherall J.C. 1998,
\apj, 506, 341
\end{thebibliography}
\end{document}